\documentclass[preprint2]{aastex}

\shorttitle{Comment on ``Uncertainty in measurements of distance''}
\shortauthors{Ng and van Dam}

\begin{document}

\title{Comment on ``Uncertainty in measurements of distance''}

\author{Y. Jack Ng and H. van Dam}
\affil{
Institute of Field Physics, Department of Physics and Astronomy,\\
University of North Carolina,
Chapel Hill, NC 27599
}
\email{yjng@physics.unc.edu}

\begin{abstract}
We have argued that quantum mechanics and general relativity give a lower
bound $\delta l \gtrsim l^{1/3} l_P^{2/3}$ on the
measurement uncertainty of any distance
$l$ much greater than the Planck length $l_P$.
Recently Baez and Olson have claimed that one can go below this bound by
attaching the measuring device to a massive elastic rod.  Here we refute
their claim.  We also reiterate (and invite our critics to
ponder on) the intimate relationship and
consistency between black hole physics (including the holographic
principle) and our bound on distance measurements.

\bigskip


\end{abstract}



We begin by recapitulating our results on distance
measurements. \citep{ngy94,ngy95}  Our
measuring device consists of a clock (which also serves as a light-emitter
and receiver) of mass $m$ and a mirror, placed respectively at the two
points the distance between which we want to measure.  By sending a light
signal from the clock to the mirror in a timing experiment, we can
determine the distance $l$.  Following Wigner\citep{wig57,sal58} we argue
that
quantum mechanics implies an uncertainty in the distance measurement
given by
\begin{equation}
\delta l \gtrsim \left(\frac{\hbar l}{mc}\right)^{1/2}.
\label{wigner}
\end{equation}
We also argue that general relativity can be used to yield
\begin{equation}
\delta l \gtrsim \frac{G m}{c^2}.
\label{nvd}
\end{equation}
Squaring the uncertainty from Eq.~(\ref{wigner}) and multiplying the result
by Eq.~(\ref{nvd}), we obtain\citep{ngy94,ngy95}
\begin{equation}
\delta l \gtrsim (l l_P^2)^{1/3},
\label{deltal}
\end{equation}
where $l_P \equiv (\hbar G/c^3)^{1/2}$ is the Planck length.\footnote{See
also the discussions in Ref.\citep{kar66} and Ref.\citep{sas99}.}

Recently Baez and Olson\citep{bae02} have argued that one can go below the
bound given above by attaching the clock (of negligible mass) to one end of
a
massive rod.  As in our own gedanken experiment, a distance $l$
is measured by
sending a burst of light from the clock and measuring the time it takes for
the light to return.  According to Baez and Olson, the uncertainty
$\delta l$ receives contributions from two sources.  They
argue that the uncertainty
of the clock's position with respect to the rod's center of mass
contributes an amount $(L \hbar / Mc)^{1/2}$, where L is the equilibrium
length of the rod and $M$ is its mass.  They also estimate that the
uncertainty of the position of the rod's center of mass contributes an
amount $(l \hbar /Mc)^{1/2}$.  Assuming that the two sources of
uncertainties are uncorrelated and arguing that the rod must be longer
than its Schwarzschild radius, i.e., $L \gtrsim GM/c^2$, they obtain
\begin{equation}
\delta l \gtrsim \left(\sqrt{\frac{GM}{c^2}} + \sqrt{l} \right)
\sqrt{\frac{\hbar}{Mc}}.
\label{BO1}
\end{equation}
By using a very heavy rod so that $l \lesssim GM/c^2$ they conclude that
\begin{equation}
\delta l \gtrsim l_P.
\label{BO2}
\end{equation}

Without commenting on the details of their argument we merely note that
their conclusion Eq.~(\ref{BO2}) depends on the inequalities
\begin{equation}
L \gtrsim GM/c^2 \gtrsim l.
\label{inequal}
\end{equation}
While one cannot expect a measuring device to be able to measure an
abitrarily small distance, one does think a good device should be able to
measure any distance bigger than a certain minimum length.  If so, then
it follows that Baez and Olson's measuring device has to be very large and
very massive.  We do not believe it can serve as an ideal clock to uncover
fundamental properties of spacetime.  The use of a huge and
massive clock will completely overwhelm the minute uncertainty in distance
measurements.  We also observe \citep{ngy00a}
that one can actually obtain Baez and
Olson's result Eq.~(\ref{BO2}) by using our own measuring device
and following our own argument if, for
the
bound on the mass $m$ of the clock, instead of Eq.~(\ref{nvd}), one uses
\begin{equation}
l \gtrsim Gm/c^2,
\label{nvd2}
\end{equation}
which is nothing but the mathematical statement of the obvious fact that,
to measure the distance from the clock to the mirror, the mirror should
not be inside the Schwarzschild radius of the clock.  Now let us recall
that Baez and Olson's measuring device has a clock attached to a massive
rod.  Thus one can regard the mass $M$ of the rod to be the effective mass
of the whole measuring device which serves as a clock.  It is curious that,
to arrive at Eq.~(\ref{BO2}), one needs the distance $l$ to be smaller
than the Schwarzshild size of the measuring device according to Baez and
Olson, whereas the opposite is true according to us.
Alternatively one can interpret Baez and Olson's disagreement with us
as arising from disagreement on whether the more restrictive bound
on $m$ given by Eq.~(\ref{nvd}) (as compared to the bound given by
Eq.~(\ref{nvd2})) is also correct.  We think so and have commented on this
issue in Ref.\citep{ngy00a}.

We do not claim that our measuring device is ideal and it is probably
impossible to give a theoretical proof that our argument is sound beyond
doubt.  In lieu of such a proof let us recall some pieces of
plausible ``circumstantial'' evidence in support of our claim.
First of all, our bound on the
uncertainty of distance measurements appears to be
consistent\citep{ngy00a,ngy00b,ngy01} with the
holographic principle\citep{tho93,sus95} which states that the maximum
number of degrees of freedom that can be put into a region of space is
given by the area of the region in Planck units.  To see this, let us
consider a region measuring $l \times l \times l$.  According
to conventional wisdom (Eq.~(\ref{BO2})), the
region can be partitioned into cubes as small
as $l_P^3$.  It follows that the number of degrees of freedom of the region
is bounded by $(l/l_P)^3$, i.e., the volume of the region in Planck units,
contradicting the holographic principle.  But according to our bound
Eq.~(\ref{deltal}), the smallest cubes inside that region have a linear
dimension of order $(l l_P^2)^{1/3}$.  Accordingly, the number of
degrees of freedom of the region is bounded by $[l/(l l_P^2)^{1/3}]^3$,
i.e., the area of the region in Planck units, as required by the
holographic principle.

It is interesting that an argument, very similar to that used by us to
derive the lower bound on the uncertainty of distance meaurements, can be
applied to relate the precision of any clock to its
lifetime. \citep{ngy01,ngy02}  For a clock
of mass $m$, if the smallest time
interval that it is capable of resolving is $t$ and its total running
time is $T$, one finds
\begin{equation}
t \gtrsim \left(\frac{\hbar T}{mc^2}\right)^{1/2},
\label{yjn1}
\end{equation}
the analogue of Eq.~(\ref{wigner}), and
\begin{equation}
t \gtrsim \frac{Gm}{c^3},
\label{yjn2}
\end{equation}
the analogue of Eq.~(\ref{nvd}).  Let us now apply these two
(in-)equalities to a black hole (of mass $m$) used as a clock.  It is
reasonable to use the light travel time across the black hole's horizon as
the resolution time\citep{bar96,ngy01} of the clock,\footnote{We
should rebut a possible objection.  One might think that, due to Hawking
radiation, the light signal cannot return to the point it started, thereby
making our ``experimental'' arrangement impossible.  But any realistic photon
detector has a finite size and it can be moved slightly in anticipation of
the return of the photon signal to detect the photon.}
i.e., $t \sim \frac{Gm}{c^3}$, then using Eq.~(\ref{yjn1})
and Eq.~(\ref{yjn2}), one immediately
finds that
\begin{equation}
T \sim \frac{G^2 m^3}{\hbar c^4},
\label{yjn3}
\end{equation}
which is Hawking's black hole lifetime!  Thus, if we had not known of black
hole evaporation, this remarkable result would have implied that there is
a maximum lifetime (of this magnitude) for a black hole.  This is another
``circumstantial'' evidence in support of our bound Eq.~(\ref{deltal})
(actually, also separately, of Eq.~(\ref{wigner}) and
Eq.~(\ref{nvd})).

There is yet another piece of ``circumstantial''
evidence.  It is related to black hole entropies and the ultimate
physical limits to computation.  But the evidence is more indirect.
Interested readers are referred to Ref.\citep{ngy02}.

Baez and Olson are not our only critics.  Some critics claim that
Eq.~(\ref{wigner}) is wrong,\citep{adl00,oza02} and some are sure that
the use of
Eq.~(\ref{nvd}) is unwarranted, though the above ``derivation'' of
Hawking's black hole lifetime seems to indicate that
there is some validity to the analogues of
the two equations.
Some critics tend to think that the relationship
between the bound on length uncertainty and the holographic principle
has not been satisfactorily proven.  Some doubt
that the above ``derivation'' of Hawking's
black hole lifetime is anything but a simple exercise in
dimensional analysis.  Conveniently they ignore the fact
that there are more than one dimensionful quantity in the problem.
Still some believe that metric
fluctuations corresponding to Eq.~(\ref{deltal}) yield an unacceptably
large fluctuations in energy density.\citep{dio96}  But actually the
energy density is extremely small\citep{ngy97} and is of the
same order of
magnitude as that for the metric fluctuations corresponding to
Eq.~(\ref{BO2}); spacetime fluctuations hardly cost any energy!

Conventional wisdom says that gravitational effects are important only
at distances comparable to the Planck scale.  Any proposals suggesting
that the uncertainty in length is not the Planck length $l_P$
should be closely scrutinized.  We think proponents of such proposals
should address the following questions:\\
1. Does the proposal contradict logic or experimental facts?\\
2. Are there hints that such a proposal deserves looking into?\\
3. What are the consequences of the proposal?\\
We have applied the above three criteria to our own proposal:\\
1. To the
best of our knowledge, our proposal does not contradict logic or
experimental facts.\\
2. That the uncertainty in length depends on more than one length scale
(as given by Eq.~(\ref{deltal})) is not surprising if we
recall\citep{ngy95} that the uncertainty in length of a thin long ruler
also depends on more than one length scale, viz., the length of the ruler
itself as well as the lattice spacing and the thermal
wavelength at low and high temperature respectively.\\
3. Our proposal is consistent with (semi-classical) black hole physics.
And the surprisingly large uncertainties arising from
distance measurements according to our proposal may one day be
detectable with improved modern gravitational-wave
interferometers\citep{ame99,ngy00b} or with improved modern laser-based
atom interferometers\citep{ngy02}.  Our work\citep{ngy97,ngy00a} also
indicates that weak gravitational waves, as needed to provide our
lower bound on uncertainty in length, do not have the energy to deform
rulers or bars
(as required for a positive signal of gravitational wave
in the Weber aluminum bar experiment), but they do deform spacetime
enough to produce the $(l l_P^2)^{1/3}$ result.\\
Our proposal adequately satisfies the three criteria.

Our arguement may not be air-tight, but we
do not think that Baez and Olson\citep{bae02} have disproved our claim.

\acknowledgments

This work was supported in part by the US Department of Energy
and by
the Bahnson Fund of the University of North Carolina.
YJN thanks J. Kowalski-Glikman for a useful email correspondence.

\end{document}